\documentstyle[11pt,aaspp4]{article}
\begin{document}
\newcommand{\msun}{M_{\odot}}
\newcommand{\zsun}{Z_{\odot}}
\newcommand{\kms}{\, {\rm Km\, s}^{-1}}
\newcommand{\cm}{\, {\rm cm}}
\newcommand{\gm}{\, {\rm g}}
\newcommand{\mum}{\, \mu{\rm m}}
\newcommand{\erg}{\, {\rm erg}}
\newcommand{\mpc}{\, {\rm Mpc}}
\newcommand{\seg}{\, {\rm s}}
\newcommand{\kev}{\, {\rm keV}}
\newcommand{\angs}{\, {\rm \AA}}
\newcommand{\hz}{\, {\rm Hz}}
\newcommand{\hi}{H\thinspace I\ }
\newcommand{\hii}{H\thinspace II\ }
\newcommand{\heii}{He\thinspace II\ }
\newcommand{\heiii}{He\thinspace III\ }
\newcommand{\nhi}{N_{HI}}
\newcommand{\lya}{Ly$\alpha$ }
\newcommand{\etal}{et al.\ }
\newcommand{\yr}{\, {\rm yr}}
\newcommand{\eq}{eq.\ }
\def\arcsec{''\hskip-3pt .}

\title{Small-Angle Scattering of X-Rays from Extragalactic Sources by Dust in
Intervening Galaxies}
\author{Jordi Miralda-Escud\'e$^{1}$}
\affil{University of Pennsylvania, Dept. of Physics and Astronomy,
David Rittenhouse Lab.,
209 S. 33rd St., Philadelphia, PA 19104}
\authoremail{jordi@llull.physics.upenn.edu}
\affil{$^{1}$ Alfred P. Sloan Fellow}

\begin{abstract}

Gamma-ray bursts are now known to be a cosmological population of objects,
which are often accompanied by X-ray and optical afterglows. The total energy
emitted in the afterglow can be similar to the energy radiated in the gamma-ray
burst itself. If a galaxy containing a large column density of dust is near the
line of sight to a gamma-ray burst, small-angle scattering of the X-rays due to
diffraction by the dust grains will give rise to an X-ray echo of the
afterglow. A measurement of the angular size of the echo at a certain time
after the afterglow is observed yields a combination of the angular diameter
distances to the scattering galaxy and the gamma-ray burst that can be used to
constrain cosmological models in the same way as a time delay in a
gravitational lens. The scattering galaxy will generally cause gravitational
lensing as well, and this should modify the shape of the X-ray echo from a
circular ring.

  The main difficulty in detecting this phenomenon is the very low flux
expected for the echo. The flux can be increased when the gamma-ray burst is
highly magnified by gravitational lensing, or when the deflecting galaxy is at
low redshift.
X-ray echos of continuous (but variable) sources, such as quasars, may
also be detectable with high-resolution instruments and would allow similar
measurements.

\end{abstract}

\keywords{ dust - gamma rays: bursts - gravitational lensing -
X-rays: general}

\section{Introduction}

  The presence of dust grains along the line of sight to an X-ray point
source will result in a halo of scattered X-rays around the source,
which will be time delayed relative to the X-rays that were not
scattered. The effect is caused by small-angle scattering of the X-rays
by the dust grains, due to the diffraction of the wavefront when some of
the X-rays are absorbed by the dust grain. The process has been
discussed previously by several authors (Overbeck 1965; Slysh 1969;
Hayakawa 1970; Tr\"umper \& Sch\"onfelder 1973; Alcock \& Hatchett 1978).
The applications that were considered were generally
restricted to galactic sources. The first observations of the phenomenon
were made by the Einstein observatory (Mauche and Gorenstein 1986 and
references therein).

  In the case of an extragalactic X-ray source with a large column
density of dust on the line of sight, the dust is likely to be
concentrated on an intervening galaxy. Therefore, all the scattering
takes place on a single screen, implying that the time delay is uniquely
determined by the angular separation from the point source. Thus, a
variable X-ray quasar located behing a dusty galaxy should show a
variable intensity pattern in the surrounding X-ray halo, with the
radial dependence reflecting the past lightcurve of the quasar. If the
time delay at a given angular separation from the source can be
measured, a combination of the angular diameter distances to the
scattering galaxy and the X-ray source is obtained, using a method
analogous to the measurement of angular diameter distances from
time delays in gravitational lensing (Refsdal 1966). This method of
measuring distances was discussed by Tr\"umper \& Sch\"onfelder for
galactic sources; in this case, the dust is likely to be distributed
along the line of sight, and  the past lightcurve of the X-ray source
should be smeared out in the X-ray halo depending on the dust
distribution.

  Recently, gamma-ray bursts have been proved to be of cosmological
origin, and the associated X-ray and optical afterglows have been
discovered (van Paradijs \etal 1997; Djorgovski \etal 1997; Metzger
\etal 1997; Costa \etal 1997).
The presence of the afterglow was predicted from
the generic model of interaction of a relativistic fireball with a
surrounding medium (Vietri 1997; M\'esz\'aros \& Rees 1997; Wijers,
Rees, \& M\'esz\'aros 1997).
Gamma-ray bursts may prove to be a useful source for doing this
type of observation.

  The paper is organized as follows. In \S 2 we derive the general
characteristics of the scattered X-rays: the angular size of the halo,
the time delay, and the intensity. In general, gravitational lensing
effects are likely to be present, since a dusty galaxy on the line
of sight will typically cause a gravitational deflection of the same
order as the scattering angle of the X-rays; this is analyzed in
\S 3. Our results are discussed in \S 4.

\section{Characteristics of the X-ray Echo}

  We consider an X-ray source from which, in the absence of scattering,
a total fluence (or energy per unit area) $\epsilon_0$ is received
instataneously; i.e., the flux from the source can be written as
$F=\epsilon_0\, \delta(t-t_0)$, where $t_0$ is the time when the
non-scattered X-rays arrive. This may be a good approximation for a
gamma-ray burst afterglow.
The source is at redshift $z_s$ and angular diameter
distance $D_s$, and the dusty galaxy is at redshift $z_d$ and angular
diameter distance $D_d$. Any scattered X-rays received at an angle
$\theta$ from the source will arrive with a time delay $t_d$ given
by:
\begin{equation}
c\, t_d = { (1+z_d)\, D_d\, D_s\over D_{ds} } ~ { \theta^2 \over 2} ~,
\label{delay}
\end{equation}
where $D_{ds}$ is the angular diameter distance from the deflector
to the source. This equation is generally valid for any cosmological
model, and is derived in the same way as the equation for the time delay
in gravitational lensing (Refsdal 1966).

  Thus, the ``echo'' of an instantaneous pulse of X-rays consists of
a ring of angular size $\theta$ at time $t_d$ after the pulse has been
received.

\subsection{Intensity of the X-ray Echo}

  To calculate the intensity of the X-ray echo, we need to know the
scattering cross section as a function of the scattering angle $\alpha$.
We consider spherical dust grains of radius $a$. The typical scattering
angle of X-rays of wavelength $\lambda$ is $\alpha \sim
\lambda/(2\pi a)$; therefore, X-rays will generally be scattered at very
small angles by grains with sizes $a \sim 1\mu m$. The total scattering
cross section of a dust grain is given by $\sigma_T = Q_s \pi a^2$,
where $Q_s$ is the scattering efficiency. For the case of magnesium
silicate, a typical material in dust grains, the scattering efficiency
can be approximated as (see van de Hulst 1957, Alcock \&
Hatchett 1978)
\begin{eqnarray}
Q_s & \simeq & 0.7 \left( {\lambda\over 2 \angs } \right)^2 \,
\left( a\over 1\mum \right)^2 \, , ~~ {\rm if}~ Q_s < 1 ~; \nonumber \\
Q_s & \simeq & 1.5\, , ~~  {\rm otherwise} .
\end{eqnarray}
The transition between the two regimes occurs when the phase shift
across the dust grain, $ | \rho | = (4\pi a)/\lambda \, | m-1 |$ (where
$m$ is the refractive index of the dust), reaches unity. For magnesium
silicate, $| \rho | \simeq ( \lambda / 2\angs ) \, ( a/ 1\mum ) $. The
detailed form of $Q_s$ is shown in Figure 1 of Alcock \& Hatchett (the
numbers in the horizontal top and bottom axes in this figure are
erroneous; see Figs. 32 and 33 in van de Hulst 1957 for similar curves).

  The differential cross section can be written as
\begin{equation}
{ d\sigma\over d\alpha} = \sigma_T\, {2\pi a \over \lambda} ~
\sigma_n \!\left( { 2\pi a \over \lambda}\, \alpha \right) ~.
\end{equation}
The dimensionless function $\sigma_n$ is normalized to $\int_0^{\infty}
dx\, \sigma_n(x) = 1$. This expression is valid in the limit of small
scattering angles (this is why the upper limit in the integral
normalizing $\sigma_n$ is replaced by infinity). The function $\sigma_n$
can be calculated for spherical grains under different approximations,
depending on the value of $ | \rho | $. The result is shown in Figure 2
of Alcock \& Hatchett; generally, $\sigma_n(x)$ rises linearly with $x$
for $x \ll 1$ (as required by continuity of the scattering cross section
per unit solid angle for zero deflection), peaks near $x\simeq 1.5$, and
then declines rapidly and may have oscillations of decreasing amplitude
at large $x$ (analogous to the familiar diffraction pattern of a circular
aperture). Practically all the X-rays are scattered at angles with
$x<4$. Of course, dust grains should not be exactly spherical in
practice, and there should be a wide distribution of grain sizes.

  Considering now a variable X-ray source with an unabsorbed flux
as a function of time $F_0(t)$, and assuming that the fraction of X-rays
scattered in the intervening galaxy is small (so that multiple
scatterings can be neglected), the surface brightness $S$ of the X-ray
echo at angular separation $\theta$ and position angle $\phi$ from the
source is
\begin{equation}
S(\theta,\phi; t)\, \sin\theta\, d\theta\, d\phi = F_0(t-t_d)\,
\tau(\theta,\phi) \,
\sigma_T^{-1} {d\sigma\over d\alpha } \, d\alpha\, (d\phi/2\pi) ~.
\end{equation}
This equation expresses the fact that the flux received along a
direction within $(d\theta, d\phi)$ is equal to the total flux of
the source, times the probability of scattering (equal to
$\tau(\theta,\phi)$, which is assumed to be small), times the
probability that the photon is scattered in the direction within
$(d\alpha, d\phi)$.
Expressing the scattering angle in terms of the angular separation,
$\alpha = \theta\, (D_s/D_{ds})$, we have
\begin{equation}
S(\theta,\phi; t) = {F_0(t-t_d)\over \sin\theta } ~
\tau(\theta,\phi)~ {a\, D_s\over \lambda \, D_{ds}} ~
\sigma_n\!\left( {2\pi a\, D_s \over \lambda\, D_{ds}} \theta \right) ~.
\end{equation}
When the variable source emits most of its energy on a time much
shorter than $t_d$ (i.e., the flux is approximated as $F_0(t)= \epsilon_0\,
\delta(t-t_0)$ as before), the echo is a narrow ring with total flux $F_e$
given by
\begin{equation}
F_e(t) = {\epsilon_0\over t_d } \, {a D_s\over 2\lambda D_{ds} } \, 
\theta ~ \sigma_n\!\left( {2\pi a\, D_s\over \lambda\, D_{ds} } \theta
\right)\, \int_0^{2\pi} d\phi\, \tau(\theta,\phi) ~,
\label{fe1}
\end{equation}
where the angle $\theta$ is to be calculated from the time delay
$t_d = t-t_0$ from equation (\ref{delay}).

  The optical depth for X-ray scattering is due to dust grains of
different sizes, and can be expressed as $\tau = \int N_d(a)\,
\sigma_T(a)\, da$, where $N_d(a)\, da$ is the column density of dust
grains of radius $a$. The dust column density can be written as $N_d(a)
= N_b\, Z_d(a)\, (4\pi a^3 \rho_d/3 m_p)^{-1}$, where $N_b$ is the
column density of baryons in the interstellar medium of the intervening
galaxy, $Z_d(a)\, da$ is the fraction of mass of the interstellar matter
in the form of grains of radius $a$ (i.e., the ``metallicity'' in dust
grains of radius $a$), $\rho_d$ is the dust density, and $m_p$ is the
proton mass. We assume that $Z_d(a)$ is independent of position.
Substituting into equation (\ref{fe1}), we obtain
\begin{equation}
F_e(t) = {\epsilon_0\over t_d } \, {3 m_p D_s\over 8 \rho_d \lambda
D_{ds} }\, \theta\, \int_0^{2\pi} d\phi \, N_b(\theta,\phi) ~ \int da \,
Z_d(a)\, Q_s(\lambda, a) ~ \sigma_n\! \left( {2\pi a D_s \over \lambda
D_{ds} }\, \theta \right) ~.
\label{fe2}
\end{equation}

  In order to proceed further, we will need to discuss the model for
the function $Z_d(a)$. Studies of interstellar reddening indicate that
$Z_d(a)\propto a^{-0.5}$ for small grains, $a \lesssim 0.1 \mum$,
(Mathis, Rumpl, \& Nordsieck 1977). The function must turn over at
larger size so that the total mass in grains converges. While most of
the mass seems to be in grains of size near $0.1 \mum$ (Martin 1978;
Savage \& Mathis 1979; Mitsuda \etal 1990), it seems natural that the
mass is distributed over a wide range of $a$. Thus, as an example,
we shall assume $Z_d(a)\propto a^{-1.5}$ for $a > 0.1 \mum$. The
scattering of X-rays in the regime we shall discuss is dominated by
grains with $a \gtrsim 1 \mum$, so our estimate of the flux is
sensitive to the uncertain form of $Z_d(a)$, which may vary depending
on the interstellar environment of the galaxy. For the adopted model,
if $Z_t$ is the total metallicity in dust grains, we have
$a Z_d(a) = Z_t/4 \, (a/0.1 \mum)^{-0.5}$. A chemically evolved galaxy
can be reasonably assumed to have $Z_t\sim Z_{\odot}$.

  Let us now examine carefully the terms appearing in the integral over
the dust grain radius in equation \ref{fe2}.
The function $\sigma_n$ is sharply peaked at the radius $a_p$ given by
$2\pi a_p D_s \theta/ (\lambda D_{ds}) \simeq 1.5$
(see Fig. 2 of Alcock \& Hatchett), and is normalized to
$\int \sigma_n(x) \, dx = 1$.
As long as $a > 1\mum\, (3\angs/\lambda) \equiv a_1$, we can use
$Q_s \simeq 1.5$ (see the discussion above; this value for $Q_s$ is an
average over the range $1 \lesssim a/a_1 \lesssim 10$, which will be the
range of interest). Smaller grains are not effective for X-ray
scattering because $Q_s$ is very small. For the assumed shape
$Z_d(a) \propto a^{-1.5}$, and if $a_p > a_1$, a sufficient
approximation is obtained by substituting $Z_d(a)$ in equation \ref{fe2}
by its value at the peak of the function $\sigma_n$.
Thus, when $a_p > a_1$, the integral can be approximated
as
\begin{equation}
\int da \, Z_d(a)\, Q_s(\lambda, a) ~
\sigma_n\! \left( {2\pi a D_s \over \lambda D_{ds} }\, \theta \right)
\simeq a_p\, Z_d(a_p) = Z_t/4 {a_p\over 0.1\mum}^{-0.5} ~.
\label{intrad}
\end{equation}
The condition $a_p > a_1$ implies
\begin{equation} 
a_p \simeq 1.5\, {\lambda D_{ds}\over 2\pi D_s \theta } > 3.8 \mum
\left( { (D_{ds}\, 1''\over D_s\, \theta } \right)^{1/2} ~,
\label{radmin}
\end{equation}
or
\begin{equation} 
\lambda > 0.78 \angs \, \left( {D_s\, \theta \over D_{ds}\, 1''}
\right)^{1/2} ~.
\label{wavmin}
\end{equation}
For shorter
wavelengths (when the angular separation $\theta$, and therefore the
time delay, is fixed), the integral rapidly becomes much smaller;
essentially, the intensity of the scattered rays is greatly reduced
because the grains that are small enough to cause scattering at an
angle $\theta$ are mostly transparent to the radiation.

  Thus, when equation (\ref{wavmin}) is satisfied, the flux of the X-ray
echo is
\begin{eqnarray}
F_e & \simeq & {\epsilon_0\over t_d}\, {6\pi m_p D_s \theta \over
8\rho_d \lambda D_{ds} } \bar N_b(\theta)\, a_p\, Z_d(a_p) \nonumber
\\ & \simeq &
0.03 {\epsilon_0\over t_d}\, {D_s\over D_{ds} }\,
{1\angs\over \lambda}\, {\theta\over 1''}\,
{\bar N_b \over 10^{22}\cm^{-2} }\, {Z_t \over 0.02} \,
{a_p \over 0.1 \mum}^{-0.5} ~, \\
\label{finalf}
\end{eqnarray}
where $\bar N_b(\theta)$ is the mean gas column density over the ring
of angular size $\theta$, and we have used the value $\rho_d =
3 \gm\cm^{-3}$ (appropriate for magnesium silicate). Notice that
$\lambda$ is the wavelength in the frame of the deflector in all our
equations (i.e., the observed wavelength is $\lambda(1+z_d)$).

  As an example, we consider a gamma-ray burst at redshift $z\sim 1$
similar to GRB970228. The total X-ray fluence of this burst in the
energy band from $2$ to $10$ keV was $4\times 10^{-6} \erg\cm^{-2}$,
about 40\% of the fluence in the gamma-ray burst itself in the band
$40-700$ keV (Costa \etal 1997; van Paradijs \etal 1997; Wijers, Rees,
\& M\'esz\'aros 1997). A large fraction of the X-ray fluence was
emitted in the beginning of the afterglow, about $50$ seconds after
the burst, and the flux decayed as $t^{-1.33}$ (Costa \etal 1997). The
X-rays that can be scattered most effectively are in the range $2$ to
$6$ keV (at lower energies, the scattering is large and the X-rays
arrive with a very long time delay, and at higher energies the
condition $a_p> a_1$ is not obeyed). We take the fluence in this
narrower energy band to be $2\times 10^{-6} \erg\cm^{-2}$.
For a time delay of one year, $\epsilon_0/t_d = 10^{-13.2} \erg\seg^{-1}
\cm^{-2}$. If the deflector is at $(1+z_d) D_d \sim 10^3 \mpc$, and
taking $D_s/D_{ds}\simeq 2$, then equation (\ref{delay}) gives $\theta
\simeq 3''$, and equations (\ref{radmin}) and (\ref{wavmin}) give
$\lambda > 2\angs$, $a_p > 1.5 \mu m$. Using the fiducial values of
$\bar N_b$ and $Z_t$ in equation (\ref{finalf}), and $\lambda = 2\angs$,
we find a flux $F_e\sim 10^{-15} \erg\cm^{-2}\seg^{-1}$. At $z_s=1$, the
corresponding luminosity is $L_e \equiv 4\pi (1+z_s)^4 D_s^2 F_e
\simeq 0.02 E_0/t_d\sim 5\times 10^{42} \erg\seg^{-1} $.
This luminosity is much larger than that of normal galaxies, so the
X-ray echo can be much brighter than any intrinsic emission from the
intervening galaxy.

  We see that the dust grains that can produce these X-ray echos are
generally of a large size compared to most of the grains, which are
responsible for interstellar reddening. Small dust grains should of
course give rise to X-ray echos of larger angular size, but these are
more difficult to observe because of the longer time delay they imply,
and because most random intervening galaxies are not larger than a few
arc seconds.

  If the X-ray echo can be resolved, and the angular radius of the ring
is measured at a time $t_d$ after the known occurrence of the
gamma-ray burst, then equation (\ref{delay}) can be used to calculate
the value $D_d D_s/D_{ds}$. Assuming that the redshifts $z_d$ and $z_s$
are known, this gives a useful constraint on the Hubble constant, as
well as the value of the radius of curvature and cosmological constant
of the universe. These can be constrained separately once the
observation is done in several sources at different redshifts. If the
X-ray echo is not resolved, one might still be able to infer the
angular size of the ring by studying the intervening galaxy in optical
and infrared light, and inferring the distribution of dust. The
lightcurve of the unresolved X-ray echo might then reveal the region
of the galaxy that the echo is moving through. For example, if the
intervening galaxy has a smooth dust density distribution peaked at
the center, but the position of the gamma-ray burst is displaced from
the center of the galaxy, then the time when the X-ray echo peaks
in intensity should correspond to the time when the echo has passed
through the center of the galaxy.

  The main difficulty in detecting the X-ray echo will be the very low
flux expected, $\sim 10^{-15} \erg\cm^{-2}\seg^{-1}$. For example, the
few X-ray clusters of galaxies detected so far at high redshift have
fluxes above $10^{-13} \erg\cm^{-2}\seg^{-1}$ (Luppino \& Gioia 1995)
(however, the X-ray echo should be detectable at a lower flux than an
X-ray cluster, because of its small angular extent). The faintest
X-ray flux detectable with the Advanced X-ray Astrophysics Facility
should be of order $10^{-15} \erg\cm^{-2}\seg^{-1}$. Other future
missions (such as XMM and Constellation X) have much larger apertures
and could reach fainter fluxes, but with lower spatial resolution, so
the photon noise is then dominated by the X-ray background for this very
low flux.

\section{Effects of Gravitational Lensing}

  The X-ray echo we have discussed should be visible when an intervening
galaxy containing large amounts of dust is close to the line of sight to
a gamma-ray burst. The gravitational field of the intervening galaxy
will very often also deflect the light-rays in a very significant way.
In fact, the typical deflection angle of gravitational lenses caused by
ordinary galaxies is of the same order as the scattering angles of the
X-ray echos we have discussed in the previous section.

  In the presence of gravitational lensing, equation (\ref{delay})
giving the time delay needs to be modified to
\begin{equation}
c\, t_d = { (1+z_d)\, D_d\, D_s\over D_{ds} }\, \left[
{ \theta^2 \over 2} - \psi(\theta,\phi)  \right] ~,
\label{gdelay}
\end{equation}
where $\phi$ is the azimuthal angle. The projected potential $\psi$
is determined by $\nabla^2\psi = 2 \kappa$, where the Laplacian
operator differentiates with respect to the angular coordinates,
$\kappa = \Sigma/\Sigma_{crit}$ is the convergence, $\Sigma$ is the
surface density of mass, and
$\Sigma_{crit} = c^2/(4\pi G)\, D_s/(D_d D_{ds})$ is the critical
surface density (e.g., Blandford \& Narayan 1992).

  If the mass of the galaxy itself were negligible, and the
gravitational lensing were caused by a smooth mass distribution around
the galaxy (due to a possible cluster of galaxies of which the
intervening dusty galaxy could be part of, or any large-scale structure
density fluctuations present along the line of sight), then we can
approximate the time delay surface with a second-order expansion. The
images of background sources are then stretched along the two orthogonal
axes of the shear by factors $(1-\kappa-\gamma)^{-1}$ and
$(1-\kappa+\gamma)^{-1}$, where $\gamma$ is the shear (determined by the
second derivatives of the projected potential). The X-ray echo would
also be stretched by the same factors, becoming an ellipse. For high
magnification, the X-ray echo increases to a given angular size in a
much shorter time than in the absence of lensing, therefore increasing
its flux through the factor $\epsilon_0/t_d$ in equation (\ref{finalf}).

  When the mass of the galaxy is important, the gamma-ray burst may be
multiply imaged. In this case, the X-ray echo should follow a curious
evolution that could in principle provide a detailed map of the
potential of the lensing mass. Considering a typical case of a
three-image lens, with the third image being near the core of a galaxy
and generally demagnified by a large factor, one should first see the
image of the gamma-ray burst in the minimum of the time delay surface.
This image would be followed by a highly elongated, arc-like X-ray echo
expanding around the image. The arc would then increase in size and
progressively bend around the center of the galaxy, until its two tips
would touch on the opposite side of the first image. The second image of
the gamma-ray burst should then arrive from the saddle-point of the time
delay surface. The X-ray echo should then break up into two rings, one
expanding out (with a shape that would continue mapping the mass
distribution around the galaxy), and the other collapsing toward the
center to eventually disappear in the third image of the gamma-ray burst
(at the maximum of the time delay surface), expected to be very faint
for the usual centrally peaked density profiles of galaxies.

  Gravitationally lensed X-ray echos should generally be brighter
than non-lensed ones, because of the shorter time it takes for the
echo to cover a given solid angle. In addition, the required
scattering angle is reduced, and this can only result in an increased
surface brightness for the echo (this increase is probably small,
because most grains are small enough to produce scattering angles
much larger than a few arc seconds). The observed time delays can
again be used to obtain the product $D_d D_s/D_{ds}$, as in other
gravitational lenses. The advantage here is that, if the evolution
of the X-ray echo can be closely monitored, the shape of the lensing
potential can be mapped, eliminating the usual modeling uncertainties
in other gravitational lenses. Notice, however, that the well-known
``$\kappa$-degeneracy'' is not eliminated: if a uniform sheet of matter
with surface density $\Sigma=\kappa_0 \,\Sigma_{crit}$ is added to the
lens, resulting in the addition of $\kappa_0\theta^2/2$ to the projected
potential, and the potential $\psi$ is multiplied by $1-\kappa_0$,
the new time delay surface differs from the initial one only by
the constant $1-\kappa_0$, which can be absorbed into the factor
$D_d D_s/D_{ds}$. Fortunately, this is the only systematic uncertainty
that remains if the mapping of the X-ray echo can yield an accurate
measurement of the shape of the time delay surface.

\section{Discussion}

  Gamma-ray bursts having a dusty galaxy near their line of sight should
be followed by an X-ray echo of the X-ray afterglow resulting from
small-angle X-ray scattering by the dust. We have discussed the expected
properties of these echos, including their angular extent as a function
of time, and their flux. The main difficulty in detecting these X-ray
echos will be their very low flux. Given the intensity of the observed
X-ray afterglow following a gamma-ray burst, and the identification of
any dust-rich galaxy near the line of sight from optical and infrared
observations, it should be possible to predict approximately the flux of
and time when the echo should be observable, and to estimate if it could
be detected.

  There are some cases where the flux of the X-ray echo of a gamma-ray
burst afterglow could be substantially higher than our estimate in \S 2.
One possibility is the presence of gravitational lensing, which we have
discussed in \S 3. Another possible case where a particularly bright
X-ray echo might be observed is when the scattering galaxy is at very
low redshift. In this case, the dust could have a larger angular extent,
and the time delay should be much shorter for a given angle. For
example, if $(1+z_d)D_d = 100 \mpc$, then the angular size is $\theta
\simeq 6''$ when the time delay is two months, and for the fluence
$\epsilon_0=2\times 10^{-6} \erg\cm^{-2}$, the flux of the echo
is now close to $10^{-14}\erg\cm^{-2}\seg^{-1}$. It needs to be pointed
out that, given the presence of intervening dust, the probability
density of the redshift of the intervening galaxy is constant at low
redshift, because the cross section of a galaxy to produce an X-ray echo
is proportional to its angular size (as opposed to the cross section for
gravitational lensing, which is constant at low redshift for lenses with
an isothermal profile).

  X-ray echos may also be searched around X-ray quasars with a dusty
galaxy near their line of sight. In this case, the fluence $\epsilon_0$
should be the integrated variable flux over some feature in the
lightcurve of the quasar, which can then be seen reverberated in the
scattered echo with a measured time delay. Of course, the detection of
the X-ray echo of a constant source would also be of great interest,
even without the measurement of the time delay; in this case, high
angular resolution is needed to separate the X-ray echo from the point
spread function of the quasar, and any X-ray emission from a possible
cluster around the quasar or intervening galaxy.

\acknowledgements

  I thank the anonymous referee for helpful comments.

%\newpage
%
%\begin{figure}
%\centerline{
%\hbox{
%\epsfxsize=4.4truein
%\epsfbox[55 32 525 706]{fig1.ps}
%}
%}
%\vskip -40pt
%\caption{Lensing diagram indicating the notation used in the text}
%\end{figure}
%\vfill\eject

\end{document}